\documentclass[showpacs,aps,pre,epsfig]{revtex4}

\usepackage{amsmath}
\usepackage{graphicx}
\usepackage{float}
\usepackage{amsfonts,dsfont}
\usepackage{amssymb}
\usepackage{color}
\usepackage{epstopdf}

\newcommand{\cP}{\mathcal{P}}
\newcommand{\cT}{\mathcal{T}}

\begin{document}

\title{Some case example 
exact solutions for quadratically nonlinear  optical  media with
 $\mathcal{PT}$-symmetric potentials}
\author{Y.N. Truong Vu}
\affiliation{ Department of Mathematics and Statistics, Amherst College, Amherst, MA, USA}
\author{J. D'Ambroise}
\affiliation{ Department of Mathematics and Statistics, Amherst College, Amherst, MA, USA}
\author{F.Kh. Abdullaev}
\affiliation{ Department of Physics, Faculty of Sciences, IIUM, Jln. Indera Mahkota, Sultan Ahmad Shah, 25200,  Kuantan, Malaysia}
\author{P.G. Kevrekidis}
\affiliation{Department of Mathematics and Statistics, University of Massachusetts,
Amherst, MA USA}

\begin{abstract}
In the present paper we  consider an optical system with a $\chi^{(2)}$-type nonlinearity and unspecified $\mathcal{PT}$-symmetric potential functions. 
Considering this as an inverse problem and positing a 
family of exact solutions in terms of cnoidal functions, 
we solve for the resulting potential functions in a way that ensures the 
potentials obey the requirements of $\mathcal{PT}$-symmetry.  We  then 
focus on case examples of soliton and periodic solutions for which 
we present a stability analysis as a function of their amplitude parameters. Finally, we numerically
explore the nonlinear dynamics of the associated waveforms to identify the
outcome of the relevant dynamical instabilities of localized and
extended states.
\end{abstract}


 \maketitle

\section{Introduction}

Recently a great deal of attention has been devoted to the investigation of quantum and classical systems with $\mathcal{PT}$-symmetric non-Hermitian Hamiltonians. It was shown in the seminal paper  \cite{Bender} that  such type of Hamiltonians can have  real eigenvalues. Due to the analogy between the Schrodinger equation and the paraxial wave equation in optics, this result has applications in optics; this is one among numerous other areas studied over the past decade. For optical beams $\mathcal{PT}$-symmetry imposes the condition on the complex refractive index $n(x)=n_r(x) + in_i(x)$: even in space for the real part of the refractive index $n_r(x) = n_r(-x)$ and odd in space for the imaginary part $n_i(x)=-n_i(-x)$. Recently effects of $\mathcal{PT}$-symmetry have been observed in  optical experiments \cite{Gao}. However, optics is certainly not the sole area where  $\mathcal{PT}$-symmetric applications have recently 
emerged. More specifically, a mechanical system realizing $\cP\cT$-symmetry
has been proposed and realized in~\cite{pt_mech}, while a major thrust of
efforts has focused on the context of electronic circuits; see e.g.
the original realization of~\cite{tsampikos_recent} and
the more recent review of this activity in~\cite{tsampikos_review}.
Additionally, we note that further intriguing realizations
of $\cP\cT$-symmetry have also emerged e.g. in the realm of
whispering-gallery microcavities~\cite{pt_whisper}.
While many of these experimental realizations have been
chiefly explored at the level of linear dynamics, the intrinsic
nonlinearity of optical systems~\cite{Gao} and the potential
nonlinearity also of electrical ones (e.g. in the form of
of a $\cP\cT$-symmetric dimer of Van-der-Pol oscillators~\cite{bend_kott})
have prompted a considerable amount of work at the interface
of nonlinearity and $\cP\cT$-symmetry.

Nonlinearity  leads to new effects in the $\mathcal{PT}$-symmetric systems, such as solitons (and vortices) in continuous~\cite{Musslimani08,achilleos} and 
discrete nonlinear 
optical media with $\cP\cT$-symmetric potentials~\cite{Dmitriev}, gap solitons in media with $\cP\cT$-symmetric periodic potentials~\cite{Aceves},  
non-reciprocity, instabilities and nonlinear $\cP\cT$-transitions
 in $\cP\cT$-symmetric (nonlinear) couplers~\cite{Ramezani,Sukhorukov,likev,pickton,barasflach,barasflach2}, 
as well as the smoothing of the spectral singularity in transmission~\cite{arxive}, among many others. 
Important applications to nonlinear plasmonic systems and metamaterials are under recent investigation~\cite{plasm,metha}. 
Additional developments  are connected with nonlinear 
$\cP\cT$-symmetric lattices \cite{Zez,KevPel,tyugin1,pelinovsky} and
$\cP\cT$-symmetry management \cite{DM,Longhi,Horne} etc. 

In nonlinear wave equations with $\mathcal{PT}$-symmetric terms,
 solitonic solutions can exist as was shown recently 
e.g. in the works 
of~\cite{Musslimani08,Abdullaev10,Abdullaev11,Konotop11,Malomed11}.  
In the case of an  NLS system with inhomogeneous in 
space loss/gain  parameters such solutions were found for linear $\mathcal{PT}$-potentials in the work \cite{Musslimani08,Abdullaev10,Salerno13,Kevrekidis13}, and for nonlinear $\mathcal{PT}$-potentials in the works \cite{Abdullaev10,Abdullaev11}, for a nonlocal NLS equation in~\cite{nonlocal} and 
for a cubic- quintic model in \cite{QC}. Naturally, it is also of interest to find exact solutions for solitons in other physically important systems.  
Recent numerical simulations of the $\chi^{(2)}$ system with $\mathcal{PT}$-symmetric potential showed the existence of stable bright solitons \cite{Moreira},
as well as of gap solitons in periodic $\cP\cT$-symmetric potentials~\cite{MKM}.
Discrete $\cP\cT$-symmetric systems with quadratic nonlinearity were also explored at the level of oligomer systems~\cite{kaidima}.

In the present work we will study  the $\chi^{(2)}$  system with $\mathcal{PT}$-symmetric potentials, describing wave processes in quadratically nonlinear media with spatially distributed gain/loss parameters. Such systems are,
in principle, of interest for nonlinear optics and potentially even for
atomic-molecular Bose-Einstein condensates trapped in complex potentials
(see, for example, a realization of imaginary potential in \cite{Oberthaler}).
In this paper, we find exact solitary and periodic solutions of the $\chi^{(2)}$ system.  We begin in Section \ref{sec:mod} by outlining the mathematical 
model.  In Section \ref{sec:cn} we derive exact solutions in terms of the 
Jacobi elliptic cnoidal function, and we present special case solutions for 
which we later examine stability properties.  
We essentially follow an inverse-function approach, somewhat 
reminiscent of~\cite{flach} to obtain such exact solutions for 
suitably tailored $\cP\cT$-symmetric potentials in the presence
of the quadratic nonlinearity.
In Section \ref{sec:stab} we 
present the stability analysis of the obtained solutions as a function
of the solution parameters (such as their amplitudes), and we show the 
results of propagation of the solutions in the (analogous to ``time") 
variable $z$.  Finally, in Section \ref{sec:conclude} we make our concluding 
statements and present a number of possibilities for future work.


\section{The model}
\label{sec:mod}

Let us consider the $\chi^2$ system describing the first harmonic (FH) and second harmonic (SH) propagation in quadratically nonlinear media with $\mathcal{PT}$-symmetric potentials as follows:
\begin{eqnarray}\label{dcgle1}
iu_z &+& d_1 u_{xx}+V_1(x)u +iW_1(x)u = u^*v \\
iv_z &+& d_2v_{xx}+ \kappa v + V_2(x)v + iW_2(x)v=u^2\notag
\end{eqnarray}
where $V_{1,2}(x)$ are even functions of $x$ corresponding to real parts of the refraction index, and $W_{1,2}(x)$ are odd functions of $x$ pertaining to 
imaginary parts thereof.  $W_{1,2}$ describe the inhomogeneous in space gain/loss.  Seeking standing waves in the form: $u(x,z) = U(x)e^{-i\omega z}$ and $v(x,z)=V(x)e^{-2i\omega z}$ we obtain the system
\begin{eqnarray}\label{dcgle2}
\omega U &+& d_1 U_{xx}+V_1(x)U +i W_1(x)U = U^*V \\
\sigma V &+& d_2 V_{xx} + V_2(x)V + i W_2(x)V=U^2 \notag
\end{eqnarray}
where $\sigma = 2\omega + \kappa$. It is useful to introduce 
the amplitude-phase decomposition in the form 
$U(x) = \rho_1(x)e^{i\theta(x)}, \ V(x)=\rho_2(x)e^{2i\theta(x)}$.
Solitonic solutions for $V=W=0$ have been 
reported e.g. in~\cite{KS,WD94(2),WD94,MST94,Hayata}, cnoidal wave solutions in \cite{Parker}, and solitons in the conservative 2D $\chi^{(2)}$ system with a potential $V\neq 0, W=0$ were considered recently in \cite{SM12}. For a review of solitary wave dynamics in quadratic systems see e.g.~\cite{buryak}.

 Assuming that $\rho_2, \theta$ are real, and that $\rho_1$ is either real or purely imaginary, we obtain the system
\begin{eqnarray}\label{dcgle3}
 \rho_1^*\rho_2 &=& \omega \rho_1  + d_1 \rho_{1,xx} -d_1\rho_1 (\theta_{x})^2 + V_1(x)\rho_1 \\
\rho_1^2 &=& \sigma \rho_2  + d_2 \rho_{2,xx} -4d_2\rho_2 (\theta_{x})^2 + V_2(x)\rho_2 \notag\\
 W_j(x)\rho_j^2  &=&  - j d_j(\rho_j^2 \theta_{x})_x   \notag
\end{eqnarray}
for $j\in\{1,2\}$.

\section{Solutions in terms of the cnoidal function}
\label{sec:cn}

We begin by writing $\rho_{1,2} = F_{1,2}(y)$ and $\theta_x = G(y)$ for $y={\rm cn}(rx,k)$.  This gives the system
\begin{eqnarray}\label{WF12gen}
F_1^{*}(y)F_2(y) &=& \omega F_1(y) + d_1r^2\Gamma_1(y)-d_1F_1(y)G^2(y)+V_1(x)F_1(y)  \\
F_1^2(y) &=&\sigma F_2(y) + d_2r^2\Gamma_2(y)-4d_2F_2(y)G^2(y)+V_2(x)F_2(y)  \notag\\
W_j(x) F_j(y) & = & jrd_j {\rm dn}(rx,k){\rm sn}(rx,k)\left(2F'_j(y)G(y)+G'(y)F_j(y)\right) \notag
\end{eqnarray}
for
\begin{eqnarray}\label{Gamma}
\Gamma_j(y) &=& y(2k^2-1-2k^2y^2)F'_j(y)+(1-y^2)(k^2y^2+1-k^2)F_j''(y)
\end{eqnarray}
with $j\in \{1,2\}$ and where the primes denote differentiation with respect to $y$.  Notice that by writing \eqref{WF12gen} in terms of both $x$ and $y=
{\rm cn}(rx,k)$ we avoid restrictions on the domain which would be applicable if we composed with an inverse function.

To find exact solutions, we apply the reverse engineering approach \cite{Abdullaev10,BK}; this type of technique was applied much earlier in order
to obtain exact traveling wave solutions in dynamical lattices~\cite{flach}. 
Our general strategy is to first specify the form of the functions $F_{1,2}(y), G(y)$ and then use \eqref{WF12gen} to solve for appropriate potentials ${V}_{1,2}(x), {W}_{1,2}(x)$. Thus, we rewrite the above equations as:
\begin{eqnarray}\label{VW}
V_1(x)&=&\dfrac{F_1^*(y)}{F_1(y)}F_2(y)-d_1r^2\dfrac{\Gamma_1(y)}{F_1(y)}+d_1G^2(y)-\omega \\
V_2(x)&=&\dfrac{F_1^2(y)-d_2r^2\Gamma_2(y)}{F_2(y)}+4d_2G^2(y)-\sigma \notag \\
W_j(x) &=& jrd_j {\rm dn}(rx,k) {\rm sn}(rx,k)\left(2\dfrac{F'_j(y)G(y)}{F_j(y)}+G'(y)\right) \notag
\end{eqnarray}
for $y={\rm cn}(rx,k)$ and $j\in \{1,2\}$. In each of the following subsections we make specific choices of $F_{1,2}$ and $G$ in such a way that the resulting potentials $V_{1,2}$, $W_{1,2}$ in \eqref{VW} do not contain terms with denominators (that may lead to singularities) and they also obey the requirements of $\mathcal{PT}$-symmetry ($V_{1,2}$ even functions of $x$ and $W_{1,2}$ odd functions of $x$).  In other words, we require at least that the conditions
\begin{eqnarray}
F_1(y) &\mid& \Gamma_1(y) \label{con1} \\
F_2(y) &\mid& \left(F_1^2(y)-d_2r^2\Gamma_2(y)\right) \label{con2} \\
F_j(y) &\mid& F'_j(y)G(y) \label{con3}
\end{eqnarray}
for $j=1,2$ are met for any choices of $F_{1,2}, G$ that we specify.

\subsection{Polynomial Functions}
\label{sec:poly}

Consider $F_{1,2}(y)$ in the form of generalized polynomials in $y={\rm cn}(rx,k)$
\begin{equation}\label{poly}
F_1(y)=i^{\{0,1\}}\sum\limits_{n=s_1}^{k_1} C_ny^{n}, \qquad F_2(y) = \sum\limits_{m=s_2}^{k_2}D_{m}y^{m}
\end{equation}
with coefficients $C_n,D_m\in\mathds{R}$, integer indexing bounds $s_1,s_2,k_1,k_2\geq 0$ with $k_1 > s_1, k_2>s_2$, and $C_{s_1}, C_{k_1}, D_{s_2}, D_{k_2}\neq 0$.  Notice that we restrict our attention here to polynomials with at least two terms.  The case of a monomial type solution will be included in the next section where we consider a more general class of power functions.  Recall that, in the derivation of \eqref{dcgle3}, $\rho_1$ is required to be either real or purely imaginary.  To show this in \eqref{poly} we have included an optional multiple of $i$ in the definition of $F_1$.  In the following subsections we outline the process of solving for $V_{1,2}, W_{1,2}$.  We separate into two cases which are convenient based on the resulting maximal power of the polynomial conditions \eqref{con1}-\eqref{con2}.

\subsubsection{Cnoidal parameter $k\neq 0$}

To proceed in solving for ${V}_1(x)$ using \eqref{VW} and assuming \eqref{poly} we must satisfy condition \eqref{con1}.  That is, we must have that $F_1(y)$ is a factor of the polynomial $\Gamma_1(y)$.  For $k \neq 0$, $\Gamma_1(y)$ will have maximal power $k_1+2$ so that \eqref{con1} amounts to the condition
\begin{equation}\label{Matchpoly1}
\sum\limits_{n={s_1-2}}^{{k_1}+2} \left((n+2)(n+1)(1-k^2)C_{n+2}+n^2(2k^2-1)C_n-(n-2)(n-1)k^2C_{n-2}\right)y^{n}= (\alpha_1y^2+\beta_1y+\gamma_1)\sum\limits_{n={s_1}}^{k_1} C_ny^{n}
\end{equation}
for some $\alpha_1,\beta_1, \gamma_1\in\mathds{R}$ with $\alpha_1 \neq 0$.  For convenience we use the convention that $C_j = 0$ for any index $j\not\in\{s_1, \dots, k_1\}$. Equating the coefficients of \eqref{Matchpoly1} then gives
\begin{eqnarray}\label{polycoeff}
&&\left( {n}^2(2k^2-1)-{\gamma_1} \right)C_n = \left(\alpha_1+({n}-2)({n}-1)k^2\right)C_{n-2}+\beta_1C_{n-1}-(n+2)(n+1)(1-k^2)C_{n+2} \\
&& \alpha_1  = -{k_1}({k_1}+1)k^2, \qquad \qquad  \qquad s_1(s_1-1)(1-k^2)=0, \qquad  \qquad (s_1+1)s_1(1-k^2)C_{s_1+1}=0 \notag
\end{eqnarray}
where the first equation is a recursion relation that holds for $n \in \{s_1,\dots, {k_1}+1\}$ and the latter three equations are obtained from equating the coefficients of the $y^{k_1+2}$, $y^{s_1-2}, y^{s_1-1}$ terms in \eqref{Matchpoly1}, respectively.  Note that the latter equations have made use of the conditions $C_{k_1}, C_{s_1} \neq 0$ and $C_{j}=0$ for $j\not\in\{s_1, \dots, k_1\}$.

From the first equation of the latter three in \eqref{polycoeff}, we now have that the coefficient $\alpha_1$ is determined by the highest degree chosen for $F_1$. The latter two equations in \eqref{polycoeff} then give us a starting point for finding more specific solutions.  That is, we can look for solutions with $k=1$ in terms of ${\rm cn}(rx,1)={\rm sech}(rx)$ and these latter two equations are satisfied.  Alternatively, we can look for solutions with $k\neq 1$ in which case the latter equations of \eqref{polycoeff} give that either $s_1=0$ so that the first term in the polynomial $F_1$ is required to be a constant, or 
the first term is required to be of degree $s_1=1$ with coefficient $C_{s_1+1}=C_2=0$.  We will proceed here to outline the general solution for cnoidal parameter $k\neq 0$.  Later towards the end of this section we will focus primarily on $k=1$ for the special case where $F_{1,2}$ are quadratic polynomials.
	
The recursive equations in \eqref{polycoeff} give us $k_1 - s_1 + 2$ conditions for the $k_1 - s_1 + 3$ constants $\{\beta_1, \gamma_1, C_{s_1},C_{s_1 +1},...,C_{k_1}\}$.   Later we will choose $\beta_1$ and then use conditions \eqref{polycoeff} to solve for the coefficients of $F_1$ and also $\gamma_1$.  Now that \eqref{con1} is satisfied by imposing \eqref{polycoeff}, we have ${V}_1(x)$ via \eqref{VW} as
\begin{equation}\label{polyV1}
{V}_1(x)= \pm F_2(y)-d_1r^2(\alpha_1y^2+\beta_1y+\gamma_1)+d_1G^2(y)-\omega
\end{equation}
with $y={\rm cn}(rx,k)$ as usual.  The plus sign in \eqref{polyV1} applies to $F_1$ real (using $i^0=1$ in \eqref{poly}) and the minus sign applies to $F_1$ purely complex (using $i^1=i$ in \eqref{poly}).  
Since the cnoidal function is an even function of $x$, $V_1(x)$ is an even function of $x$ so that this potential is compatible with $\mathcal{PT}$-symmetry.  Notice that $V_1(x)$ may not be a polynomial if $G^2$ is not a polynomial.  In Section \ref{sec:stab}, we will specify choices for the $G$ function and we will choose $\omega$ so that $V_1 \rightarrow 0$ as $x\rightarrow \infty$.

To solve for $V_2(x)$ we must have that $F_2(y)$ statisfies condition \eqref{con2}.  One way to proceed is to require that
\begin{equation}\label{con2F2}
F_1^2(y) = F_2(y)P(y)
\end{equation}
where $P(y)$ is a polynomial in $y$.  Then, similar to the $F_1$ case, 
we may also impose that $F_2(y) \mid \Gamma_2(y)$ so that $\Gamma_2(y) = (\alpha_2y^2+\beta_2y+\gamma_2)F_2(y)$ for some constants $\alpha_2,\beta_2,\gamma_2\in\mathds{R}$ with $\alpha_2\neq 0$.  Using a similar procedure as in the $F_1$ case, now equations \eqref{polycoeff} must hold after performing the replacements $n \rightarrow m$, $C \rightarrow D$ and in the subscripts $1 \rightarrow 2$.  Using this $F_2$-version of equation \eqref{polycoeff}, now the coefficient $\alpha_2$ is determined by the highest degree of the polynomial $F_2$.  Since $k\neq 1$ here the $F_2$-version of the latter two equations in \eqref{polycoeff} either requires us to take $s_2=0$ so that the first term in the polynomial $F_2$ is required to be a constant, or alternately the first term is required to be of degree $s_2=1$ with coefficient $C_{s_2+1}=C_2=0$.   Also, the $k_2-s_2+3$ constants $\{\beta_2, \gamma_2, D_{s_2},D_{s_2 +1},...,D_{k_2}\}$  are required to satisfy the same recursive $k_2-s_2+2$ equations in \eqref{polycoeff} but with appropriate $F_2$-version described above.

The most obvious choice in order to satisfy both \eqref{con2F2} and the $F_2$-version of \eqref{polycoeff} is to take $F_2(y), F_1(y)$ as scalar multiples
of each other.  In other words,
\begin{equation}\label{F12multiples}
F_2(y) = i^{\{0,1\}}A F_1(y)\mbox{ \  and \  } P(y) = \frac{F_1(y)}{i^{\{0,1\}}A} 
\end{equation}
for some $A\in\mathds{R}_{\neq 0}$.  Since $\rho_2, F_2$ are required to be real-valued the multiple of $i$ in front is included only if it's included in the definition of $F_1$ in \eqref{poly}.  Now we have the real-valued potential function
\begin{equation}\label{polyV2}
V_2(x)=\frac{F_1(y)}{i^{\{0,1\}}A}-d_2r^2(\alpha_2 y^2+\beta_2 y+\gamma_2)+4d_2G^2(y)-\sigma
\end{equation}
where later in specific examples we will choose $\sigma$ to be such that $V_2\rightarrow 0$ as $x\rightarrow \infty$.

Next we want to determine an appropriate form for the function $\theta_x=G(y)\in\mathds{R}$ with $y={\rm cn}(rx,k)$ that will satisfy \eqref{con3}.  Since $F_1, F_2$ have been chosen to be scalar multiples of each other, if \eqref{con3} holds for $j=1$ then it holds for $j=2$.  So, we take
\begin{equation}\label{chsG}
G(y) = T(y)F_1(y)
\end{equation}
for a function $T(y)$  and this gives via \eqref{VW}
\begin{eqnarray}\label{polyW12}
W_j(x) &=& jrd_j {\rm dn}(rx,k) {\rm sn}(rx,k)\left(2F'_1(y)T(y)+G'(y)\right)
\end{eqnarray}
for $y={\rm cn}(rx,k)$ and $j\in\{1,2\}$.  Since ${\rm sn}(rx,k)$ is an odd function of $x$ and ${\rm dn}(rx,k)$ is even, $W_1, W_2$ are odd functions of $x$ as required by $\mathcal{PT}$-symmetry as long as the quantity $2F'_1({\rm cn}(rx,k))T({\rm cn}(rx,k))+G'({\rm cn}(rx,k))$ is an even function of $x$.    This is reasonable since cn$(rx,k)$ is an even function of $x$.

Now we have a complete solution of \eqref{WF12gen} given by $F_{1,2}$ in \eqref{poly}, $V_{1,2}$ in \eqref{polyV1} and \eqref{polyV2}, and $W_{1,2}$ in \eqref{polyW12}, all under the conditions seen in \eqref{polycoeff}, \eqref{F12multiples}, \eqref{chsG}.  To be more explicit, let us focus on details in the case where $F_1$ is a quadratic function and $k=1$ so that $y={\rm sech}(rx)$.  Consider $F_1$ of the form $F_1 = C_0+C_1y+C_2y^2$ for $C_0,C_2 \neq 0$.  Then $s_1=0$ and $k_1 = 2$ so that the latter two equations in \eqref{polycoeff} are satisfied.  Now the remaining equations in \eqref{polycoeff} give $\alpha_1 = -6$, $\gamma_1 = 0$ and the conditions
\begin{equation}\label{sysdeg2}
C_1 = \beta_1C_0,
\qquad
C_2 = \dfrac{-6C_0+\beta_1C_1}{4},
\qquad
C_3 = 0 = \dfrac{(\alpha_1+2)C_1+\beta_1C_2}{9},
\end{equation}
that we may solve for the four constants $\beta_1, C_0, C_1, C_2$.

In the case that $\beta_1=0$, \eqref{sysdeg2} gives $C_1=0$ and $C_2=-{3C_0}/{2}$ so that combining with \eqref{F12multiples} we have
\begin{equation}\label{f1deg2}
F_1(y) = C_0(1-\dfrac{3}{2}y^2),
\qquad
F_2(y)=AF_1(y).
\end{equation}

Proceeding with $G(y)$ as in \eqref{chsG} for any function $T(y)$ we have by \eqref{polyV1}, \eqref{polyV2} and \eqref{polyW12} that
\begin{eqnarray} \label{quadbeta=0}
V_1(x) &=& AC_0 (1-\dfrac{3}{2}y^2)+6d_1r^2y^2+d_1G^2(y)-\omega   \\
V_2(x) &=& \dfrac{C_0}{A} (1-\dfrac{3}{2}y^2)+6d_2r^2y^2+4d_2G^2(y)-\sigma \notag\\
W_j(x) &=& jrd_j {\rm tanh}(rx) {\rm sech}(rx)\left(-6C_0yT(y)+G'(y)\right).\notag
\end{eqnarray}
Equations \eqref{f1deg2}-\eqref{quadbeta=0} give us a solution that
we will refer to as  the {\it dark-dark soliton} case.  In Section \ref{sec:stab} we show the dark soliton shape, analyze the stability of the dark-dark soliton, and show plots over the propagation variable $z$ for a specific choice of the function $G$ and other parameters.

We also consider here the case of $\beta_1 \neq 0$, for which \eqref{sysdeg2} gives two possibilities for the coefficients of the polynomial $F_1$.  Then combined with \eqref{F12multiples} we have
\begin{equation}\label{betapm}
F_1(y) = C_0(1 \pm \sqrt{22}y + 4y^2),
\qquad
F_2(y)=AF_1(y).
\end{equation}
Letting $G(y)$ be as in \eqref{chsG} for some $T(y)$ function we then obtain
\begin{eqnarray}\label{betapmVW}
V_1(x) &=& AC_0 (1 \pm \sqrt{22}y + 4y^2)-d_1r^2(\pm \sqrt{22}y-6y^2)+d_1G^2(y)-\omega \\
V_2(x) &=& \dfrac{C_0}{A} (1 \pm \sqrt{22}y + 4y^2)-d_2r^2(\pm \sqrt{22}y-6y^2)+4d_2G^2(y)-\sigma \notag\\
W_j(x) &=& jrd_j {\rm tanh}(rx) {\rm sech}(rx)\left(2C_0(8y \pm \sqrt{22})T(y)+G'(y)\right)\notag
\end{eqnarray}
for $j\in\{1,2\}$.
These solutions are quite interesting in their own right,
as the one with the $+$ sign corresponds to an {\it antidark-antidark}
soliton setting of a pair of bright solitary waves on top of a
non-vanishing background. On the other hand, the solution with 
the $-$ sign is especially structurally complex, resembling
a conglomeration of multiple --more specifically of 4-- dark
solitons. 

\subsubsection{Cnoidal parameter $k=0$ and $y=\cos(rx)$}

For $F_{1,2}$  of the polynomial form \eqref{poly} now consider the case of $k=0$, or $y=\cos(rx)$.    In solving for $V_1(x)$ the polynomial condition analogous to \eqref{Matchpoly1} has maximal power $k_1$.  This roughly makes sense because when we differentiate a $\cos(rx)$ or $\sin(rx)$ 
the result is a function of the same overall power (in contrast to derivatives of sech$(rx)$ and tanh$(rx)$, for example). The analogue of \eqref{Matchpoly1} in this case is
\begin{equation}\label{Matchpolyk=0}
\sum\limits_{n={s_1-2}}^{k_1} \left((n+2)(n+1)C_{n+2}-n^2C_{n}\right)y^n=\gamma_1\sum\limits_{n=s_1}^{k_1} C_n y^n
\end{equation}
for $\gamma_1\neq 0$.  Equating coefficients we get
\begin{eqnarray}\label{polyk0}
C_n &=& \dfrac{(n+2)(n+1)C_{n+2}}{{n}^2+{\gamma_1}}  \text{ \  for \ } n \in \{s_1,\dots, {k_1-1}\} \notag\\
\gamma_1 &=& -k_1^2, \qquad  \qquad s_1(s_1-1)=0, \qquad  \qquad (s_1+1)s_1C_{s_1+1}=0
\end{eqnarray}
where the latter three equations came from equating the coefficients of the $y^{k_1}$, $y^{s_1-2}, y^{s_1-1}$ terms in \eqref{Matchpolyk=0}, respectively.  As before, the first equation of the latter three in \eqref{polyk0} shows that the coefficient $\gamma_1$ is determined in terms of the maximal power $k_1$ of the polynomial $F_1$.  The latter two equations in \eqref{polyk0} then show that either $s_1=0$ so $F_1$ must have a constant term, or alternatively the first term is required to be of degree $s_1=1$ with coefficient $C_{s_1+1}=C_2=0$.  The remaining recursive equations in \eqref{polyk0} then give us $k_1-s_1$ conditions for $k_1-s_1+1$ unknowns $\{C_{s_1},\dots,C_{k_1}\}$. Choosing one of these coefficients will lead us to find the others.  In solving for $V_2$ we have similar conditions to \eqref{polyk0} for the constants $\gamma_2\neq 0$ and $\{D_{s_2}, \dots, D_{k_2}\}$ where in \eqref{polyk0} one should replace $n \rightarrow m$, $C \rightarrow D$ and in the subscripts $1 \rightarrow 2$.  We proceed in a similar way as in the $k\neq 0$ case above,
assuming the forms of $F_2, G$ as seen in \eqref{con2F2}, \eqref{F12multiples}, \eqref{chsG} and
finally we have
\begin{eqnarray} \label{polyVWk=0}
V_1(x) &=& \pm F_2(y)-d_1r^2\gamma_1+d_1G^2(y)-\omega \\
V_2(x) &=& \dfrac{F_1(y)}{i^{\{0,1\}}A}-d_2r^2\gamma_2+4d_2G^2(y)-\sigma \notag\\
W_j(x)&=& jrd_j \sin(rx)\left(2F'_1(y)T(y)+G'(y)\right)\notag
\end{eqnarray}
for  $j\in \{1,2\}$.

Let us focus on the details of the quadratic case where $F_1 = C_0+C_1y+C_2y^2$ for $C_0,C_2 \neq 0$ and $s_1=0$, $k_1 = 2$.  \eqref{polyk0} then gives $\gamma_1 = -4, C_0 = -C_{2}/2,$ and $C_1=0$.   Then, we have
\begin{equation}\label{f1deg2osc}
F_1=C_0(1-2y^2),
\qquad
F_2(y)=AF_1(y).
\end{equation}
We also have $\gamma_2 = -4$ by the $V_2$ analogue of \eqref{polyk0} (see description above).  Proceeding with $G(y)$ as in \eqref{chsG} for some function $T(y)$ we have
\begin{eqnarray} \label{quadk=0}
V_1(x) &=& AC_0 (1-2y^2)+4d_1r^2+d_1G^2(y)-\omega \\
V_2(x) &=& \dfrac{C_0}{A} (1-2y^2)+4d_2r^2+4d_2G^2(y)-\sigma \notag\\
W_j(x) &=& jrd_j \sin(rx)(-8C_0yT(y)+G'(y)) \notag
\end{eqnarray}
for  $j\in \{1,2\}$.
Equations \eqref{f1deg2osc}-\eqref{quadk=0} give us a solution that we call the {\it quadratic oscillatory} case.  In Section \ref{sec:stab} we show the shape of the solution, analyze the stability, and explore its dynamics over the
evolution variable ($z$).

\subsection{Power Functions}
\label{sec:pow}

Next we take $F_1(y)$ and $F_2(y)$ to be power functions
\begin{equation}\label{Fpow}
F_1(y) = i^{\{0,1\}}Cy^{p_1}, \qquad F_2(y)=Dy^{p_2}
\end{equation}
for  $p_1,p_2 \geq 0$ and $C, D\neq 0$.  
This special case considerably simplifies the relevant 
compatibility conditions. In particular,
substituting \eqref{Fpow} into \eqref{VW}
and examining conditions \eqref{con1}-\eqref{con2} we require that either $k=1$ and $2p_1 \geq p_2$, or $p_1=p_2=1$.  

\subsubsection{Cnoidal parameter $k=1$ and $y={\rm sech}(rx)$}

In the case of $k=1$ we can have non-integer $p_1$ and $p_2$; this is in contrast to the polynomial case.    For this case, we apply \eqref{VW} and find
\begin{eqnarray}\label{powk=1}
V_1(x) &=& \pm Dy^{p_2}-d_1r^2p_1(1-2y^2)-d_1r^2p_1(p_1-1)(1-y^2)+d_1G^2(y)-\omega \\
V_2(x) &=& \pm \dfrac{C^2}{D}y^{2p_1-{p_2}}-d_2r^2{p_2}(1-2y^2)-d_2r^2{p_2}({p_2}-1)(1-y^2)+4d_2G^2(y)-\sigma \notag \\
W_j(x) &=& j r d_j \tanh(rx)\left(2p_jG(y)+yG'(y)\right) \notag
\end{eqnarray}
for $j\in\{1,2\}$.  As long as $(2p_jG(y)+yG'(y))$ are even functions in $x$, which is easy to choose since $y={\rm sech}(rx)$ 
is an even function of $x$, then $W_1$ and $W_2$ are odd and compatible with the $\mathcal{PT}$ symmetry criterion.  We refer to the solutions given in \eqref{Fpow} and \eqref{powk=1} as the {\it bright-bright} soliton 
case.  In section \ref{sec:stab} we show the 
wave's shape, analyze its stability and explore its 
direct numerical evolution.
{Note that the case $p_1=p_2 =2$ corresponds to solitonic solutions found 
by Karamzin-Sukhorukov
in~\cite{KS}, and $2p_1=p_2=2$ to solitonic solutions found by Menyuk et al.~\cite{MST94}}.

\subsubsection{Powers $p_1={p_2}=1$}
In the case $p_1={p_2}=1$, we have
\begin{eqnarray}\label{powk=0}
V_1(x) &=& \pm Dy-d_1r^2(2k^2-1-2k^2y^2)+d_1G^2(y)-\omega \\
V_2(x) &=& \pm \dfrac{C^2}{D}y-d_2r^2(2k^2-1-2k^2y^2)+4d_2G^2(y)-\sigma \notag \\
W_j(x) &=& jrd_j{\rm dc}(rx,k){\rm sn}(rx,k)\left(2G(y)+yG'(y)\right) \notag
\end{eqnarray}
where $j\in\{1,2\}$.  Similar to the previous section, we only need to choose a function $G(y)$ so that $(2G(y)+yG'(y))$ is an even function of $x$ in order to satisfy the $\mathcal{PT}$ symmetry criterion.  Equations \eqref{Fpow} and \eqref{powk=0} give us a solution we call the {\it linear oscillatory} case.  
More details are included in Section \ref{sec:stab}.

\subsection{Other solutions}
\label{sec:other}

Here, we introduce a possibility which is distinct from the previous
ones as follows.
We introduce $F_1(y)$ and $F_2(y)$ in the form
\begin{equation}\label{complex}
F_1 = iAy^{p}(1-y^2)^{1/2}, \qquad F_2 = By^{q}
\end{equation}
for $p,q\geq 0$ i.e., a non-polynomial form.  
By \eqref{VW}
and examining conditions \eqref{con1}-\eqref{con2} we find that we must have $p(p-1)(1-k^2)=q(q-1)(1-k^2)=0$ and $q\leq 2p$.  That is, we require that either $k=1$ with $q\leq 2p$, or $p=q=1$, or $p=q=0$.
We will focus on the former two cases.  As for $W_{1,2}$ in \eqref{VW} and condition \eqref{con3}, we find that
 $G(y)$ must be in the form
\begin{equation}\label{otherG}
G(y)=Cy^a(1-y^2)^b
\end{equation}
where $a,b \in \mathds{N}$.

\subsubsection{Cnoidal parameter $k=1$ and $y={\rm sech}(rx)$}

In the $k=1$ case, we have the solutions
\begin{eqnarray}\label{complexF12solution}
\omega &= -d_1r^2{p}^2, \qquad &V_1(x) = -By^q+d_1r^2y^2(p+1)(p+2)+d_1G^2(y)\\
\sigma &= -d_2r^2q^2, \qquad &V_2(x)=-\frac{A^2}{B}y^{2{p}-{q}}(1-y^2)+d_2r^2y^2q(q+1)+4d_2G^2(y) \notag
\end{eqnarray}
with $q\leq 2p$ and $G(y)$ in the form of \eqref{otherG}.  Since for this family of solutions the form of $G$ is specified, it is immediately clear which choices of $\omega,\sigma$ will give $V_{1,2}\rightarrow 0$ as $x\rightarrow\infty$.  In contrast to previous sections, those choices have been made in \eqref{complexF12solution}.  Also, we have by \eqref{VW}
\begin{eqnarray}\label{complexWi}
W_1 &=& Crd_1 {\rm sech}(rx) \tanh(rx)\left(y^{a-1}(1-y^2)^b(a+2p)-2y^{a+1}(1-y^2)^{b-1}(b+1)\right) \notag \\
W_2 &=& 2Crd_2 \tanh(rx)\left(y^a(1-y^2)^b(a+2q)-2by^{a+2}(1-y^2)^{b-1}\right).
\end{eqnarray}
Equations \eqref{complex}-\eqref{complexWi} give us a solution that bears
a bright soliton coupled with a dark-in-bright soliton. The latter
involves a pair of bright solitary waves coupled in a bound state
anti-symmetric (i.e., they bear a phase difference of $\pi$) 
configuration; another example of this form has been previously
reported e.g. in~\cite{njp}.  
More details on the propagation of this solution and its stability are included in Section \ref{sec:stab}.

\subsubsection{Powers $p=q=1$}

In the case where $p=q=1$, we obtain
\begin{eqnarray}\label{complexV12p=q=1}
\omega&=-d_1r^2(5k^2-4), \qquad  &V_1(x)=-By+6d_1r^2k^2y^2+d_1G^2(y)\\
\sigma &= -d_2r^2(2k^2-1), \qquad &V_2(x)=-\dfrac{A^2}{B}y(1-y^2)+2d_2r^2k^2y^2+4d_2G^2(y).\notag
\end{eqnarray}
Again here since $G$ is known we have made choices of $\omega,\sigma$ reflected in \eqref{complexV12p=q=1} so that $V_{1,2}\rightarrow 0$ as $x\rightarrow\infty$.  With $G(y)$ as in \eqref{otherG} we have
\begin{eqnarray}\label{complexW12p=q=1}
W_1 &=& rCd_1 {\rm sn}(rx,k) {\rm dn}(rx,k)\left(y^{a-1}(1-y^2)^b(a+2)-2y^{a+1}(1-y^2)^{b-1}(b+1)\right) \notag \\
W_2 &=&  2rCd_2 {\rm dc}(rx,k) {\rm sn}(rx,k)\left(y^a(1-y^2)^b(a+2)-2by^{a+2}(1-y^2)^{b-1}\right).
\end{eqnarray}
We will refer to the solution in \eqref{complex}-\eqref{otherG} and \eqref{complexV12p=q=1}-\eqref{complexW12p=q=1} as the {\it non-polynomial oscillatory} 
solution.  In the special case of $k=1$, this reverts to a waveform
of the same type as the one examined above (namely, a bright solitary
wave coupled to a dark-in-bright one).
More details are provided on this solution in Section \ref{sec:stab}.

\section{Stability and Dynamics of the Solutions}
\label{sec:stab}

To study the stability of solutions we will first present 
the corresponding linear stability analysis framework.  We begin by writing
\begin{equation}\label{uv12stab}
u=(U(x) + a(x)e^{\lambda z} + b(x)^{\ast}e^{\lambda^{\ast}z} )e^{{-}i\omega z}, \qquad v = (V(x) + c(x)e^{\lambda z} + d(x)^{\ast}{e^{\lambda^{\ast}z}} )e^{{-}2i\omega z}
\end{equation}
where $U(x), V(x)$ are the exact solutions of \eqref{dcgle2} found in Section \ref{sec:cn}.  Substituting \eqref{uv12stab} into the system \eqref{dcgle1} we obtain in the first order set of equations
\begin{equation}
i\begin{pmatrix} \omega + L_1 & -V & -U^{\ast} & 0 \\
V^{\ast} & {-}\omega - L_1^{\ast} & 0 & U \\
-2U & 0 & {}2\omega + \kappa + L_2 & 0\\
0 & 2U^{\ast} & 0 & {-}2\omega -\kappa -L_2^{\ast} \end{pmatrix}
\begin{pmatrix} a \\ b \\ c \\ d \end{pmatrix} =
\lambda  \begin{pmatrix} a \\ b \\ c \\ d \end{pmatrix}, \label{matrix1}
\end{equation}
where
the operators $L_{1,2}$ are $L_i = d_i\partial_{xx} +V_i(x) + iW_i(x)$ for $i\in\{1,2\}$.  If, for a given solution, the corresponding eigenvalue $\lambda$ has 
a positive real part then the solution is unstable as is readily seen in \eqref{uv12stab}; otherwise the solution is stable.

In the following subsections, we will apply the linear stability analysis and show the results of numerical propagation of the solutions we found in Section \ref{sec:cn} according to \eqref{dcgle1} using a standard 
explicit 4th order Runge-Kutta code.  We focus 
on the three solitonic solutions derived in Section \ref{sec:cn}, of 
dark-dark, bright-bright and also bright coupled with the dark-in-bright
waveforms, and also on the three oscillatory solutions derived in 
Section \ref{sec:cn} in each of the quadratic, linear and non-polynomial
cases considered. In each subsection, we start by specifying  $G(y)$ and other parameters as is necessary.  We find that in all cases the solutions are unstable with increasing strength of instability as the amplitude parameters increase.  Each example we consider has various regions of weak and strong instability as is discussed in the following subsections.

Notice also that the equation
\begin{equation}\label{POWchk}
\frac{d}{dz}P(z) = -2\int \left( W_1(x)|u(x,z)|^2 + W_2(x)|v(x,z)|^2 \right) dx
\end{equation}
can be derived from the system \eqref{dcgle1} where the combined power function $P(z)$ is defined as $P(z) = \int\left( |u(x,z)|^2 + |v(x,z)|^2 \right) dx$.  Equation \eqref{POWchk} acts as a numerical check of all of the simulations performed in this section.

\subsection{Solitonic Solutions ($k=1$)}

\subsubsection{Dark-dark solitary wave}

\begin{figure}
\includegraphics[scale=.55]{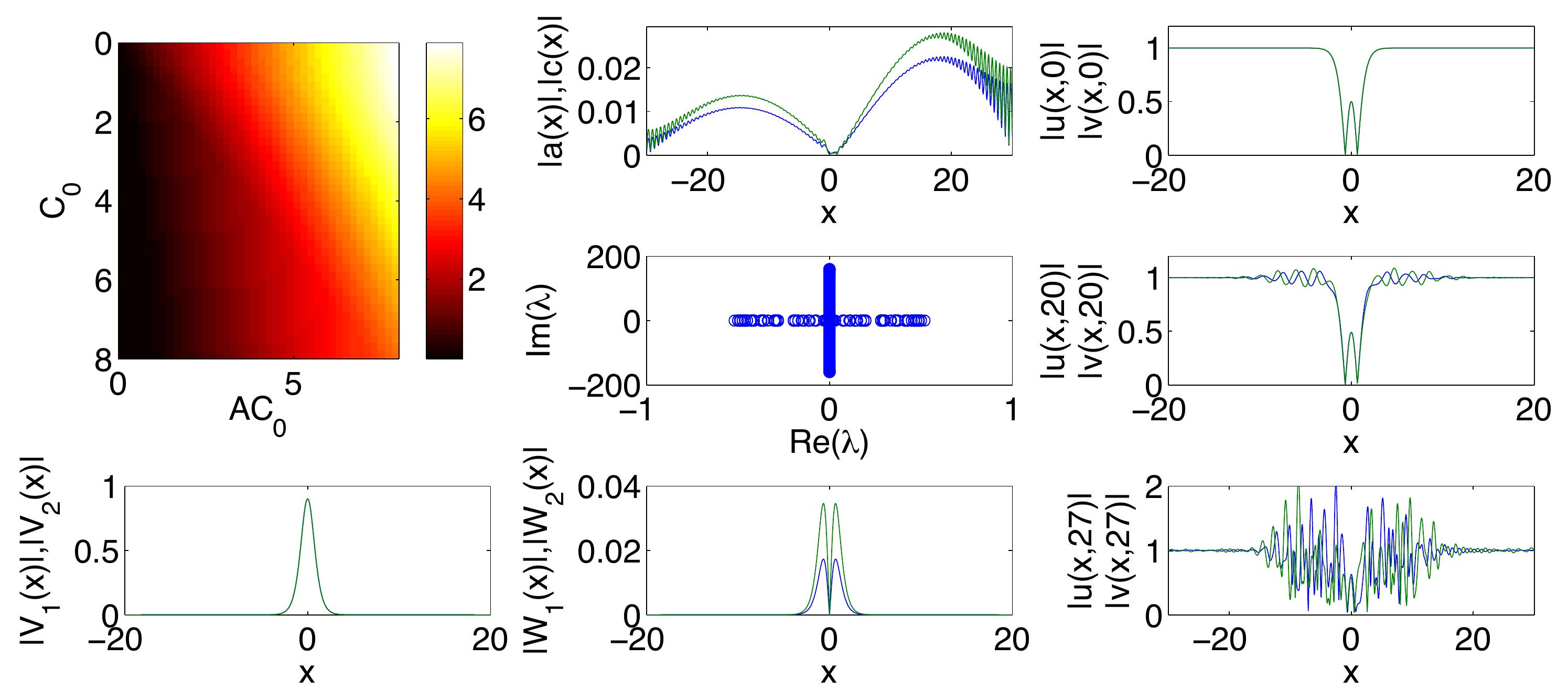}
\caption{This figure shows stability analysis and time propagation plots for the dark-dark soliton case.  The parameters $d_1=d_2=0.1$, $r=1$ and $K=0.05$ are fixed in every plot.   The contour plot in the upper left depicts ${\rm max}({\rm Re}(\lambda))$ as a function of amplitude parameters $C_0$ and $AC_0$ of the $F_1, F_2$ functions.   The other seven plots correspond to the point $(1,1)$ in the dim but non-zero region of the contour plot.  At this point we have ${\rm max}({\rm Re}(\lambda)) \approx 0.5201$.
In the left two plots of the bottom row, we show the magnitudes of the real and imaginary parts of the potential functions:  $|V_1(x)|, |V_2(x)|$ and $|W_1(x)|, |W_2(x)|$ respectively.  
In the top two plots of the center column, we show the magnitudes of the eigenvectors $|a(x)|$ (blue), $|c(x)|$ (green) and the eigenvalues $\lambda$ in the complex plane that correspond to the stationary solutions seen in the top right panel.
The right column shows the magnitudes of the solution at $t=0$ and at later times.  Here we find that the unstable solution loses its dark soliton shape over time, with the destabilization manifesting across much of the $x$ axis.
}\label{quadk=1fig}
\end{figure}

For the solutions presented in Section \ref{sec:poly} in equations \eqref{f1deg2}-\eqref{quadbeta=0} we additionally make the choice here of $G(y)=KF_1(y)$ with $K \in \mathds{R}$.  Also choosing $\omega,\sigma$ so that $V_{1,2}\rightarrow 0$ as $x\rightarrow\infty$ gives
\begin{eqnarray} \label{specificquadk=1}
\omega &= AC_0 + d_1K^2, \qquad & V_1 = AC_0 (1-\dfrac{3}{2}y^2)+6d_1r^2y^2+d_1K^2(1-\dfrac{3}{2}y^2)^2-\omega \\
 \sigma &=\dfrac{C_0}{A}+4d_2K^2, \quad &V_2 = \dfrac{C_0}{A} (1-\dfrac{3}{2}y^2)+6d_2r^2y^2+4d_2K^2(1-\dfrac{3}{2}y^2)^2-\sigma \notag\\
&&W_j = - 9jrd_jK \tanh(rx) {\rm sech}^2(rx).
\end{eqnarray}
We present the stability analysis of this family of solutions in Figure \ref{quadk=1fig}.  We find that as the amplitudes $C_0, AC_0$ of the $F_1, F_2$ functions, respectively, increase the solution becomes increasingly unstable.   The panels of time propagation plots in Figure \ref{quadk=1fig} show that over the dynamical evolution, these unstable dark soliton solutions will not 
maintain the dark soliton shape. Instead, the wide range of unstable
eigenmodes in the system will induce a form of ``lattice turbulence''
whereby the end dynamical result will appear to bear no clear
solitonic (or other) structure.

\subsubsection{Antidark-antidark solitary wave}

\begin{figure}
\includegraphics[scale=.55]{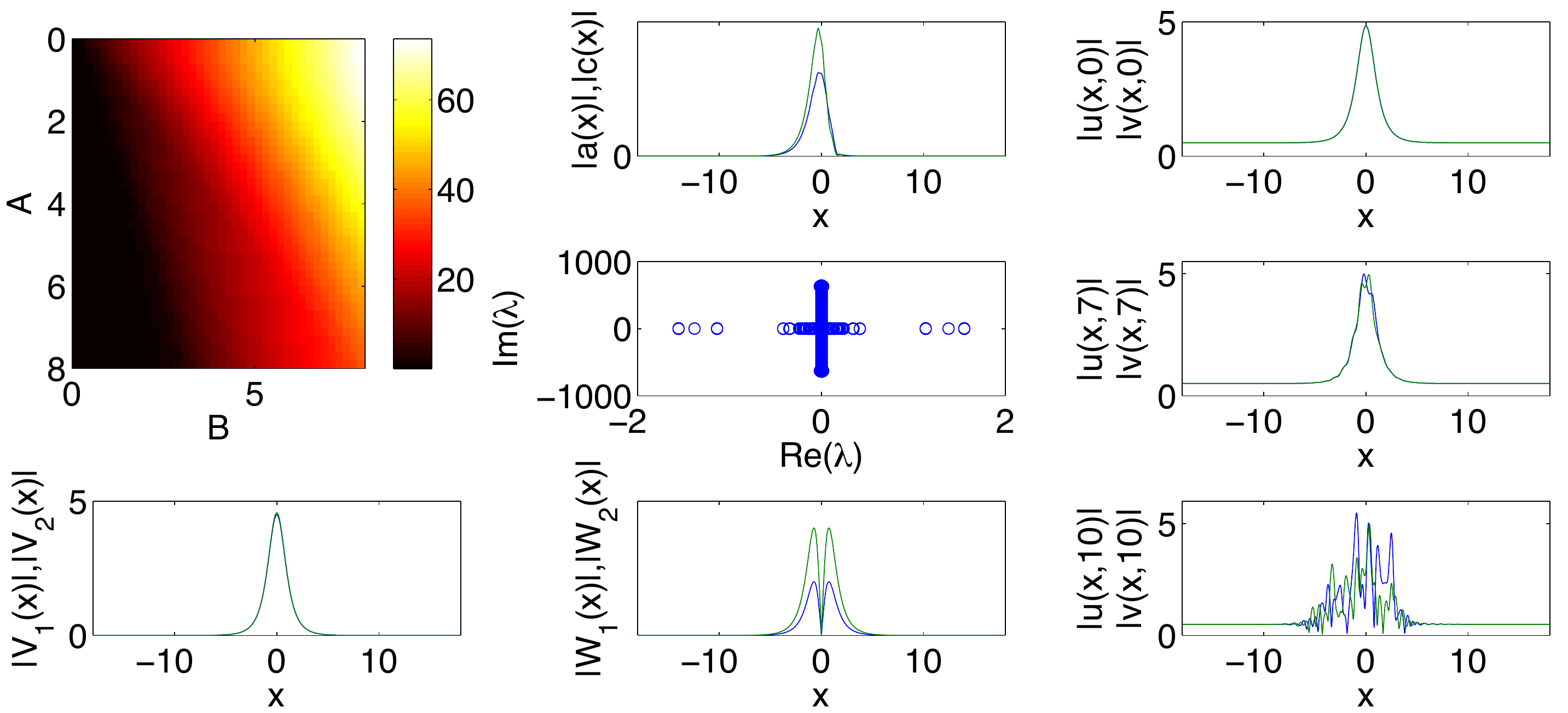}
\caption{This figure shows stability analysis and time propagation plots for the antidark-antidark soliton case.  The placement of the figures follows the same pattern as that of Figure \ref{quadk=1fig}.  The parameters $d_1=d_2=0.1$, $r=1$ and $K=0.05$ are fixed in every plot.   The point $C=D=0.5$ from the upper left contour plot corresponds to ${\rm max}({\rm Re}(\lambda)) \approx 1.5547$, and the other seven plots show details about these amplitude values.   Here we find that the unstable solution loses its soliton shape in a way that is similar to the dark-in-bright example below in Figure \ref{otherk=1fig}.
}\label{betapfig}
\end{figure}

For the solutions presented in Section \ref{sec:poly} in equations \eqref{betapm}-\eqref{betapmVW} with the $+$ sign, we take $G(y)=KF_1(y)$ with $K \in \mathds{R}$.  Also choosing $\omega,\sigma$ so that $V_{1,2}\rightarrow 0$ as $x\rightarrow\infty$ gives
\begin{eqnarray} \label{specificbetap}
\omega &= AC_0 + d_1K^2, \qquad & V_1 = AC_0 (\sqrt{22}y+4y^2)-d_1r^2(\sqrt{22}y-6y^2)+d_1K^2(1+\sqrt{22}y+4y^2)^2-d_1K^2 \\
 \sigma &=\dfrac{C_0}{A}+4d_2K^2, \quad &V_2 = \frac{C_0}{A} (\sqrt{22}y+4y^2)-d_2r^2(\sqrt{22}y-6y^2)+4d_2K^2(1+\sqrt{22}y+4y^2)^2-4d_2K^2  \notag\\
&&W_j = 3jrd_jK {\rm tanh}(rx){\rm sech}(rx) \left( \sqrt{22} + 8{\rm sech}(rx)\right).
\end{eqnarray}
We present the stability analysis of this family of solutions in Figure \ref{betapfig}.  We find that as the amplitudes $C_0, AC_0$ of the $F_1, F_2$ functions, respectively, increase the solution becomes increasingly unstable with a pattern similar to the previous example.   However, here the eigenvectors are localized.  The panels of time propagation plots in Figure \ref{betapfig} show that over the dynamical evolution, these soliton solutions will not maintain the soliton shape. Instead, the turbelence occurs near the center of the lattice, close to the solution's peak.  The instability is similar over time to what is observed in the dark-in-bright example below.

\subsubsection{Multiple dark solitary wave}

\begin{figure}
\includegraphics[scale=.55]{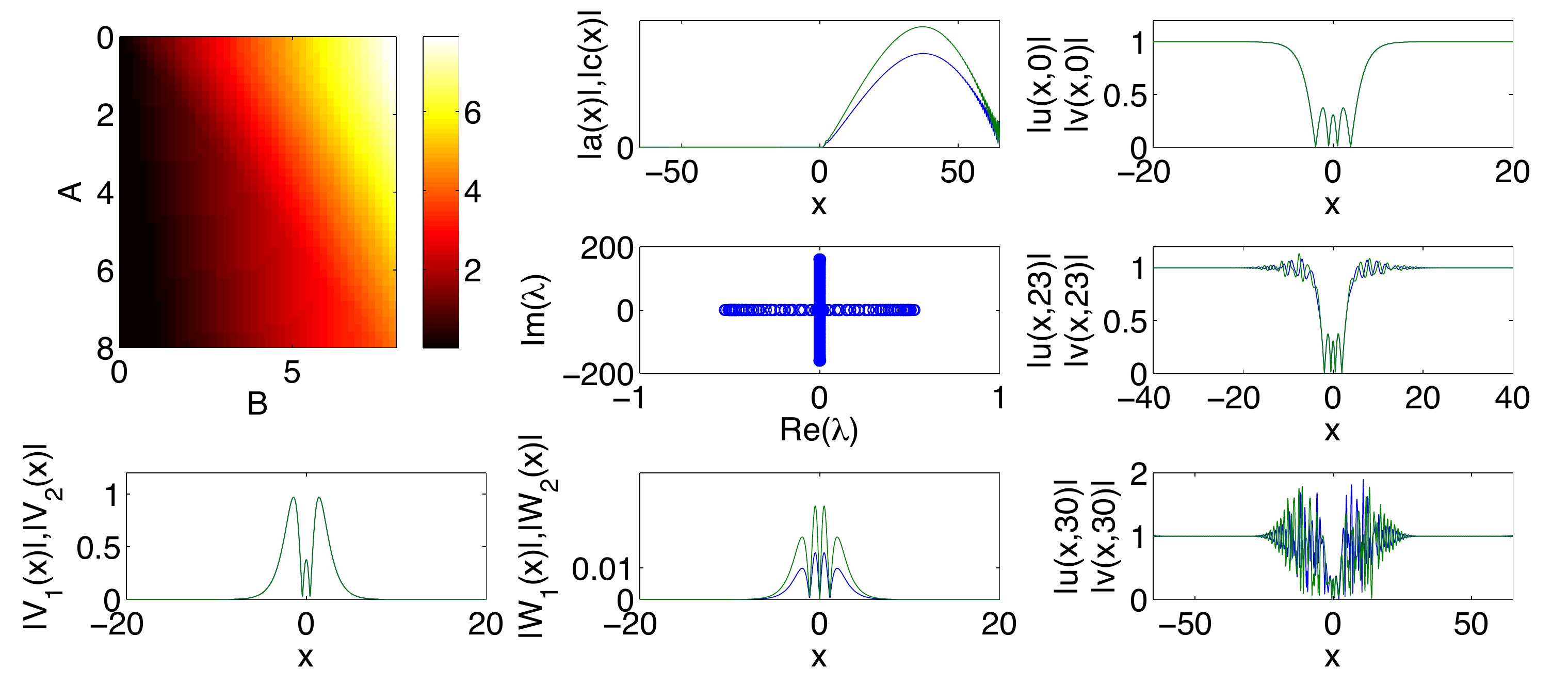}
\caption{This figure shows stability analysis and time propagation plots for the multiple dark soliton case.  The placement of the figures follows the same pattern as that of Figure \ref{quadk=1fig}.  The parameters $d_1=d_2=0.1$, $r=1$ and $K=0.05$ are fixed in every plot.   The point $C=D=0.5$ from the upper left contour plot corresponds to ${\rm max}({\rm Re}(\lambda)) \approx 0.5277$, and the other seven plots show details about these amplitude values.   Here we find that the unstable solution loses its dark soliton shape similar to the first dark-dark example.
}\label{betamfig}
\end{figure}

For the solutions presented in Section \ref{sec:poly} in equations \eqref{betapm}-\eqref{betapmVW} with the $-$ sign, we take $G(y)=KF_1(y)$ with $K \in \mathds{R}$.  Also choosing $\omega,\sigma$ so that $V_{1,2}\rightarrow 0$ as $x\rightarrow\infty$ gives
\begin{eqnarray} \label{specificbetam}
\omega &= AC_0 + d_1K^2, \qquad & V_1 = AC_0 (-\sqrt{22}y+4y^2)+d_1r^2(\sqrt{22}y+6y^2)+d_1K^2(1-\sqrt{22}y+4y^2)^2-d_1K^2 \\
 \sigma &=\dfrac{C_0}{A}+4d_2K^2, \quad &V_2 = \frac{C_0}{A} (-\sqrt{22}y+4y^2)+d_2r^2(\sqrt{22}y+6y^2)+4d_2K^2(1-\sqrt{22}y+4y^2)^2-4d_2K^2  \notag\\
&&W_j = 3jrd_jK {\rm tanh}(rx){\rm sech}(rx) \left( -\sqrt{22} + 8{\rm sech}(rx)\right).
\end{eqnarray}
We present the stability analysis of this family of solutions in Figure \ref{betamfig}.  We find that as the amplitudes $C_0, AC_0$ of the $F_1, F_2$ functions, respectively, increase the solution becomes increasingly unstable with a pattern similar to the previous example.   The panels of time propagation plots in Figure \ref{betamfig} show that over the dynamical evolution, these soliton solutions will not maintain the dark soliton shape. Instead, the turbelence occurs across the $x$ axis.  The instability is similar over time to what is observed in the first dark-dark example above.

\subsubsection{Bright-Bright solitary wave} 

\begin{figure}
\begin{center}
\includegraphics[scale=.55]{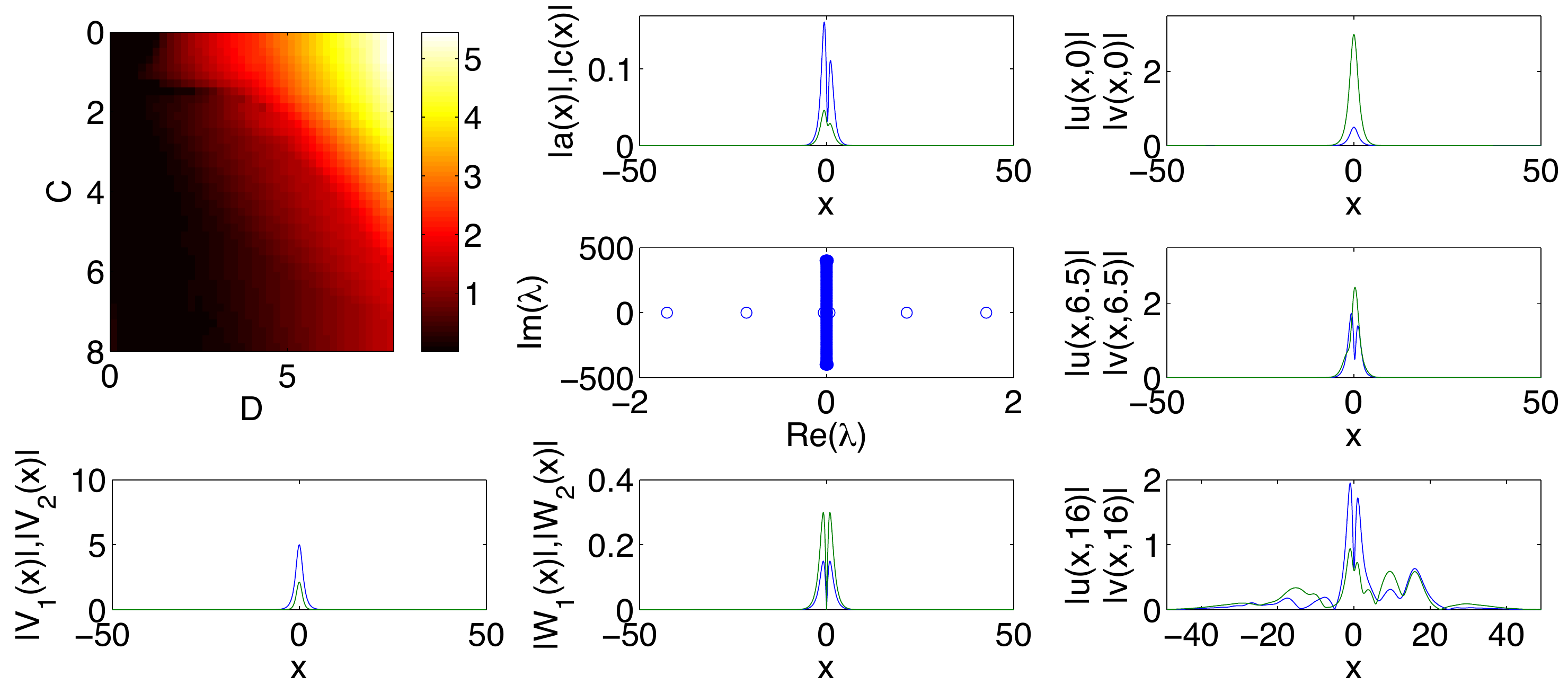}
\end{center}
\caption{This figure shows stability analysis and time propagation plots for the bright-bright soliton case  with $p=q=1$.   The placement of the figures follows the same pattern as that of Figure \ref{quadk=1fig}.  Here the parameters common to all plots are $d_1=d_2=1$, $r=1$ and $K=0.1$.  The point $C=0.5, D=3$ from the upper left contour plot corresponds to ${\rm max}({\rm Re}(\lambda)) \approx 1.7055$ and more details regarding these specific amplitude parameters are shown in the other seven plots.  Here we find that as time progresses the solution loses its soliton shape with turbulent, symmetry-breaking behaviour occurring first nearby the central peak and then leaking outward across the $x$ axis.  
}
  \label{powk=1fig}
\end{figure}

For the solutions presented in \ref{sec:pow} in equations \eqref{Fpow} and \eqref{powk=1} we simply choose $G(y)=Ky$ with $K \in \mathds{R}$ and we take $\omega,\sigma$ to be such that  $V_{1,2}\rightarrow 0$ as $x\rightarrow\infty$.  This gives the solution
\begin{eqnarray}\label{specificmonok=1}
\omega &=& -d_1r^2{p_1}^2, \quad V_1=\pm Dy^{p_2}+d_1K^2y^2+d_1r^2{p_1}({p_1}+1)y^2, \quad W_1=rd_1K(2{p_1}+1) {\rm sech}(rx) {\rm tanh}(rx) \\ \notag
\sigma &=& -d_2r^2{p_2}^2, \quad V_2=\pm \dfrac{C^2}{D}y^{2{p_1}-{p_2}}+4d_2K^2y^2+d_2r^2{p_2}({p_2}+1)y^2, \quad W_2=2rd_2K(2{p_2}+1) {\rm sech}(rx) {\rm tanh}(rx).
\end{eqnarray}
In Figure \ref{powk=1fig} we show the stability analysis for selected parameters.  Similar to the dark-dark case, the solutions do not maintain their shape as time progresses and the strength of instability increases as both the multipliers $C$ and $D$ in \eqref{Fpow} increase.  Here, it is clear that the instability
results in the breaking of the parity symmetry, leading to a symmetry-breaking
pattern. 

\subsubsection{Bright and Dark-in-Bright solitary Wave} 

\begin{figure}
\includegraphics[scale=.55]{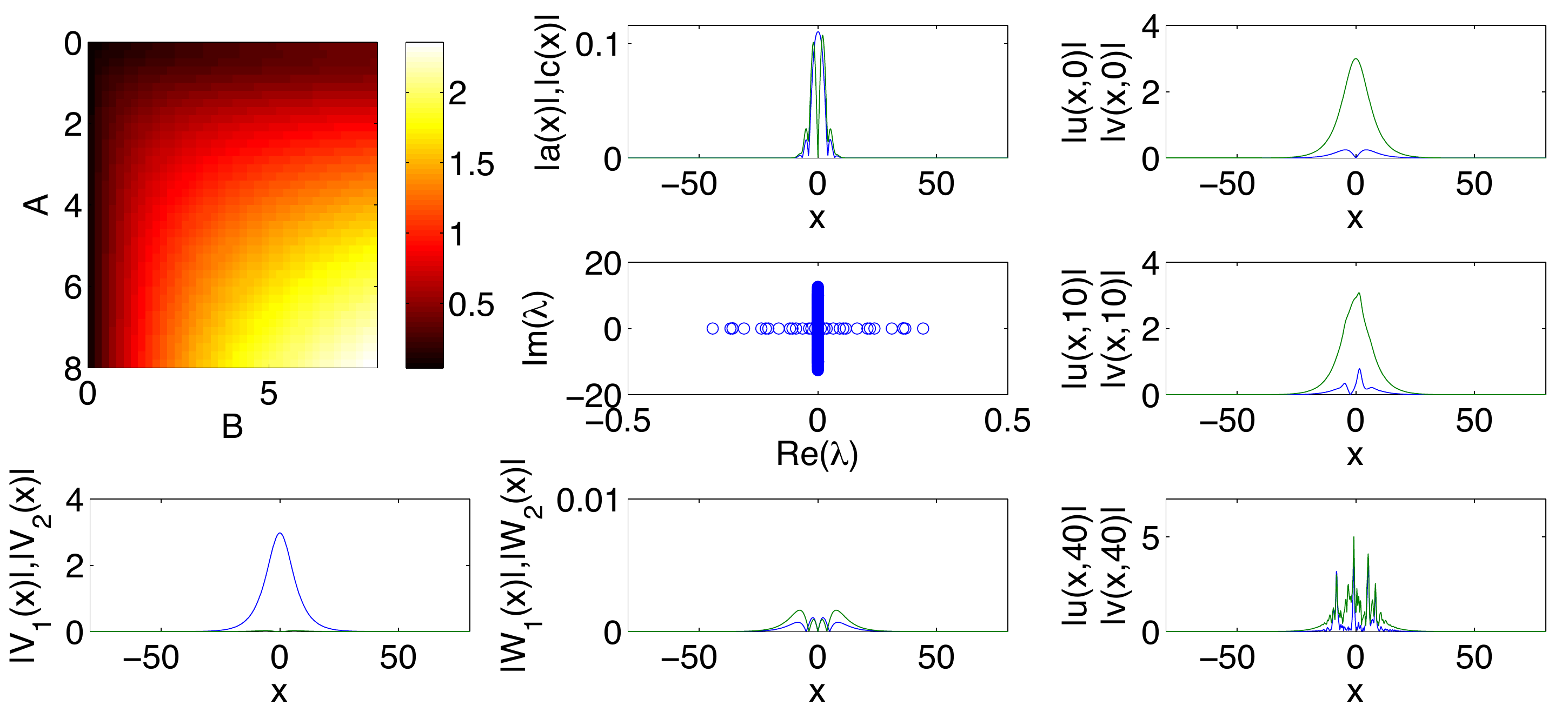}
\caption{This figure shows stability analysis and time propagation plots for the bright and dark-in-bright soliton case.  The placement of the figures follows the same pattern as that of Figure \ref{quadk=1fig}.  Here the parameters common to all plots are $d_1=d_2=0.1$, $r=0.2$ and $K=0.05$.  The point $C=0.5, D=3$ from the upper left contour plot corresponds to ${\rm max}({\rm Re}(\lambda)) \approx 0.2778$ and more details regarding these specific amplitude parameters are shown in the other seven plots.  Here we find that as time progresses the solution loses its soliton shape completely.
}\label{otherk=1fig}
\end{figure}

For the solutions presented in Section \ref{sec:other} in equations \eqref{complex}-\eqref{complexWi} we choose $G(y)=Ky(1-y^2)$ and obtain
\begin{eqnarray}\label{specificotherk=1}
\omega&=-d_1r^2p^2, \qquad &V_1=-By^q+d_1r^2y^2(p+1)(p+2)+d_1K^2y^2(1-y^2)^2 \notag \\
\sigma&=-d_2r^2q^2, \qquad &V_2=-\dfrac{A^2}{B}y^{2p-q}(1-y^2)+d_2r^2y^2q(q+1)+4d_2K^2y^2(1-y^2)^2 \notag\\
&&W_1 = Krd_1 {\rm sech}(rx) \tanh(rx)(1+2p-(5+2p)y^2) \notag \\
&&W_2= 2Krd_2 {\rm sech}(rx)\tanh(rx)(1+2q-(3+2q)y^2) \notag
\end{eqnarray}
by choosing $\omega,\sigma$ as usual.  The stability analysis is presented in Figure \ref{otherk=1fig}.  Similar to the quadratic case, the strength of instability increases as $A$ increases and as $B$ increases.  The propagation plots in Figure \ref{otherk=1fig} show that the peak destabilizes and the amplitude spreads out over the $x$ axis while maintaining some comparative concentration at the center of the axis. Furthermore, a symmetry-breaking feature appears once
again to be amplifed and be distinctly observable at the end
of the simulation's reporting horizon.

\subsection{Oscillatory Solutions ($k=0$)}

\subsubsection{Quadratic oscillatory solution}

\begin{figure} 
\includegraphics[scale=.55]{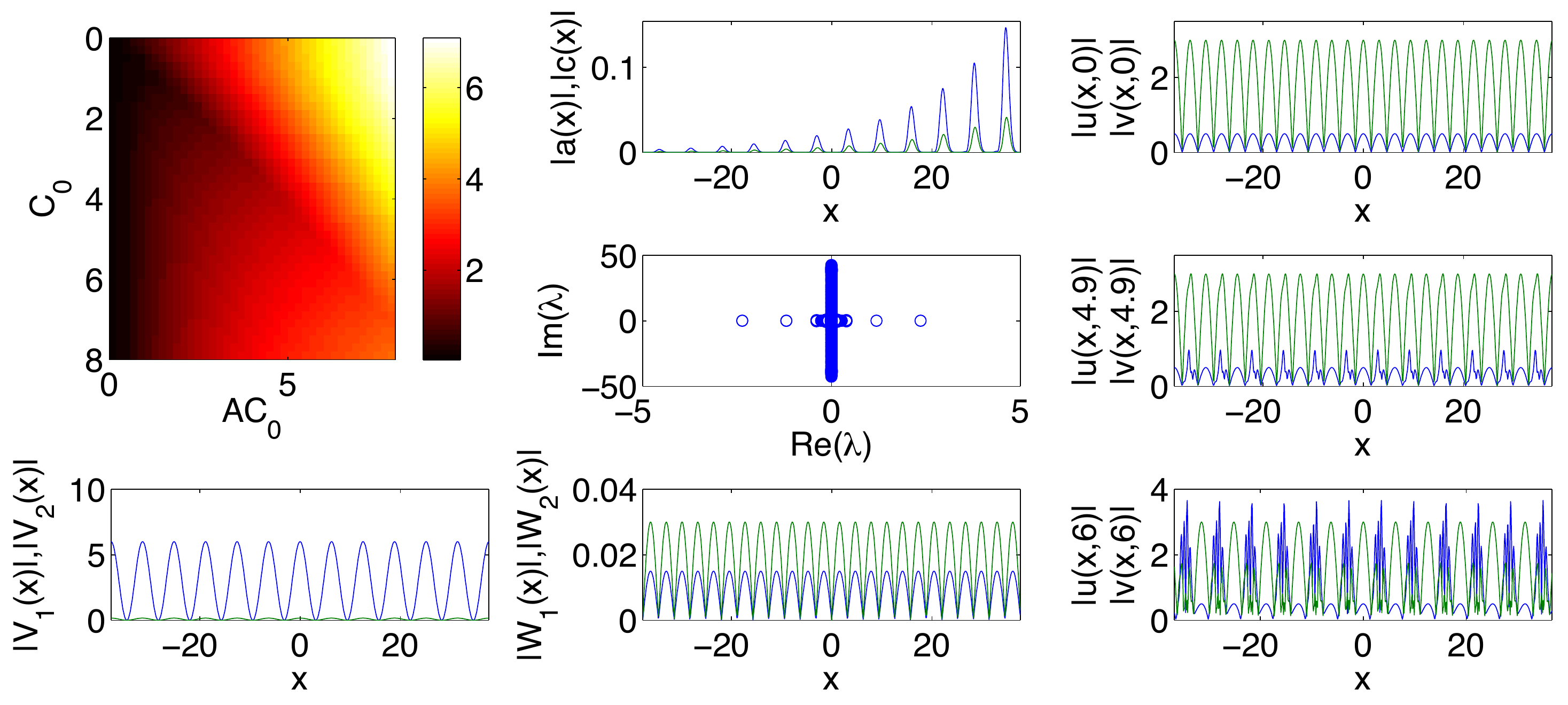}
\caption{This figure shows stability analysis and time propagation plots for the quadratic oscillatory case.   The placement of the figures follows the same pattern as that of Figure \ref{quadk=1fig}.  Here the parameters common to all plots are $d_1=d_2=0.1$, $r=0.5$ and $K=0.05$.  The point $C=0.5, D=3$ from the upper left contour plot corresponds to ${\rm max}({\rm Re}(\lambda)) \approx 2.3602$ and more details regarding these specific amplitude parameters are shown in the other seven plots.  Here we find that as time progresses the instability of the solution appears across the $x$ axis at near regular intervals, close to the periodicity of the stationary solution.  However, the solution over time does not remain truly periodic.  
}
\label{quadk=0fig}
\end{figure}

For the solutions in Section \ref{sec:poly} in equations \eqref{f1deg2osc}-\eqref{quadk=0} we take $G(y)=KF_1(y)$ similar to the dark-dark soliton case, 
obtaining the following solutions
\begin{eqnarray} \label{specificquadk=0}
\omega &= AC_0 + d_1K^2 +4d_1r^2, \qquad
&V_1 =  -2AC_0y^2+d_1K^2(1-2y^2)^2-d_1K^2 \\
\sigma &=\dfrac{C_0}{A}+4d_2K^2+4d_2r^2, \quad
&V_2 = -2\dfrac{C_0}{A} y^2+4d_2K^2(1-2y^2)^2-4d_2K^2 \notag\\
&& W_j = -12jrd_jK \cos(rx) \sin(rx) \notag
\end{eqnarray}
for $j\in\{1,2\}$.  The stability graph in Figure \ref{quadk=0fig} has similar features  to the one in Figure \ref{quadk=1fig}, showing that the changes in stability strength of the system across the amplitudes $C_0, AC_0$ grid are similar despite very different $k$ values. In the current oscillatory function case, we observe that the waves will not maintain their original shapes, with the most apparent distortions located at near-periodic points along the $x$ axis. These distortions will not only break the periodicity of the structure but they will also
lead (within some lattice periods of the solution) into the 
turbulent dynamical evolution discussed previously.

\subsubsection{Linear oscillatory solution} 

For the solutions in Section \ref{sec:pow} in equations \eqref{Fpow} and \eqref{powk=0} we take $G(y)=Ky$ and obtain the solution
\begin{eqnarray}\label{specificmonok=0}
\omega &=& d_1r^2, \quad V_1=\pm Dy+d_1K^2y^2, \quad W_1=3rd_1K \sin(rx) \\ \notag
\sigma &=& d_2r^2, \quad V_2=\pm \dfrac{C^2}{D}y+4d_2K^2y^2, \quad W_2=6rd_2K \sin(rx).
\end{eqnarray}

\begin{figure}
\includegraphics[scale=.55]{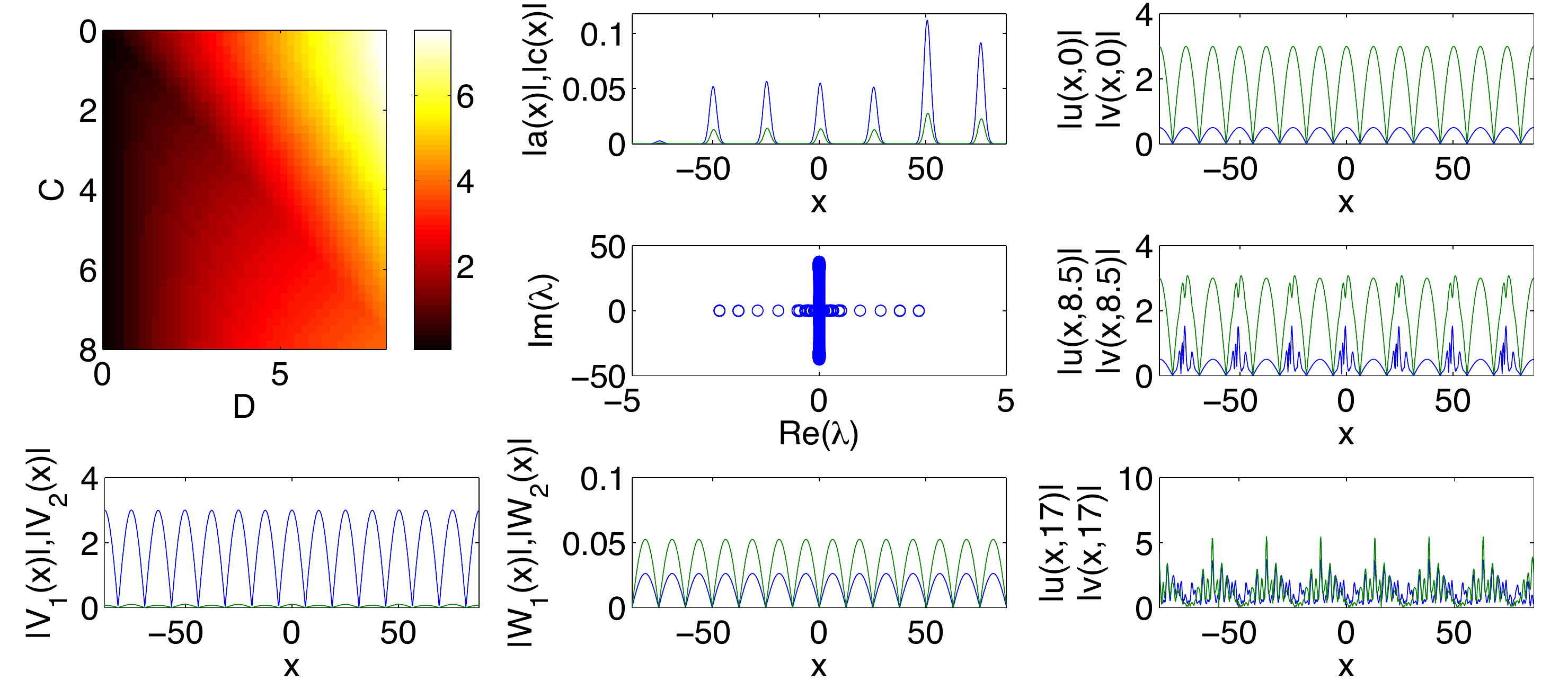}
  \caption{This figure shows stability analysis and time propagation plots for the linear oscillatory case.   The placement of the figures follows the same pattern as that of Figure \ref{quadk=1fig}.  Here the parameters common to all plots are $d_1=d_2=0.35$, $r=0.25$ and $K=0.1$.  The point $C=0.5, D=3$ from the upper left contour plot corresponds to ${\rm max}({\rm Re}(\lambda)) \approx 2.6630$ and more details regarding these specific amplitude parameters are shown in the other seven plots.  Similar to the quadratic oscillatory case, here we find that as time progresses the solution does not remain truly periodic.  
 }\label{powk=0fig}
\end{figure}

The stability graph in this linear function case for $k=0$ is presented in Figure \ref{powk=0fig}.  Here we see that the pattern of the strength of the instability is more similar to that of the quadratic functions than it is to the linear functions case with $k=1$. We can see that the strength of instability increases as $C,D$ increase.  The propagation plots in Figure \ref{powk=0fig} show that over time the wave loses its original shape at points across the $x$ axis. Here the instability is  induced by  eigenvectors which also extend across the $x$ axis and which lead to a breakup of the periodicity of the original pattern. 

\subsubsection{Other oscillatory solution} 

\begin{figure}
\includegraphics[scale=.55]{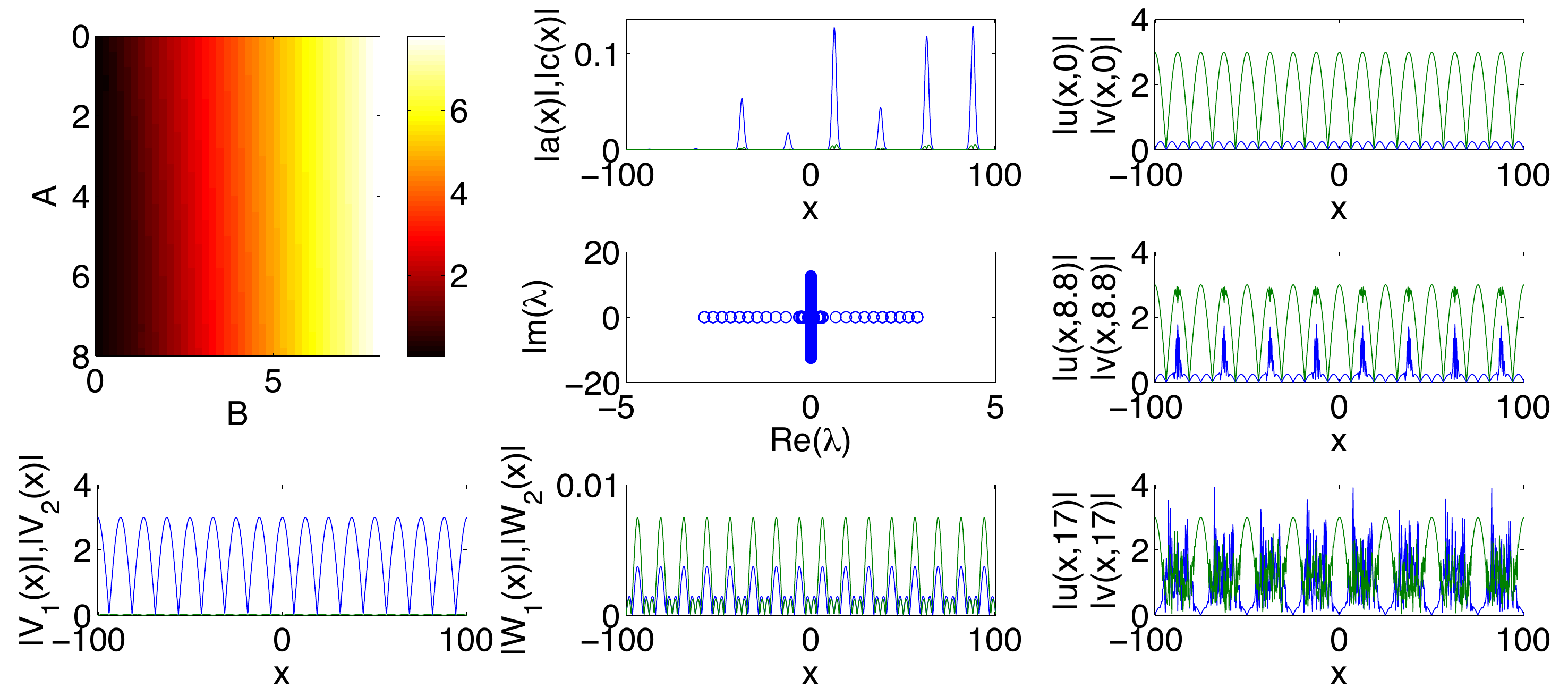}
\caption{This figure shows stability analysis and time propagation plots for the other oscillatory case.   The placement of the figures follows the same pattern as that of Figure \ref{quadk=1fig}.  Here the parameters common to all plots are $d_1=d_2=0.1$, $r=0.25$ and $K=0.05$.  The point $C=0.5, D=3$ from the upper left contour plot corresponds to ${\rm max}({\rm Re}(\lambda)) \approx 2.8819$ and more details regarding these specific amplitude parameters are shown in the other seven plots.  Here we find that as time progresses the instability of the solution appears across the $x$ axis at near regular intervals, close to the periodicity of the stationary solution.  The instability is similar to the other oscillatory cases and again leads to a periodicity breakup. 
}\label{otherk=0fig}
\end{figure}

For the solutions in Section \ref{sec:other} in equations \eqref{complex}-\eqref{otherG} and \eqref{complexV12p=q=1}-\eqref{complexW12p=q=1} we choose $G(y)=Ky(1-y^2)$, $p=q=1$, and obtain
\begin{eqnarray}\label{specificotherk=0}
\omega &= 4d_1r^2, \qquad & V_1= -By+d_1K^2y^2(1-y^2)^2 \\
\sigma &=d_2r^2, \quad &V_2= -\dfrac{A^2}{B}y(1-y^2)+4d_2K^2y^2(1-y^2)^2 \notag\\
&&W_1 = Krd_1 \sin(rx)\left(3-7\cos^2(rx)\right) \notag \\
&&W_2 = 2Krd_2 \sin(rx)\left(3-5\cos^2(rx)\right). \notag
\end{eqnarray}
The stability graph on the left  of Figure \ref{otherk=0fig} shows that the strength of instability increases as $B$ increases, yet it appears to be 
roughly independent of $A$. The propagation panels show 
the distortion of the original solution occurring over time at points across the $x$ axis corresponding to an eigenvector that is also spread across $x$.

\section{Conclusion}
\label{sec:conclude}


In the present work, we have explored both solitary and more broadly
periodic (including cnoidal and even their trigonometric limit of
$k=0$) solutions of the $\cP\cT$-symmetric problem with quadratic
nonlinearity. A reverse engineering approach was
adopted herein attempting to identify even real potentials
and odd imaginary ones that would be compatible with specific
cnoidal solutions (and their hyperbolic limits in the case of $k=1$,
as well as their trigonometric ones in the case of $k=0$).
It was shown that necessitating the existence of such solutions
generally leads to a number of plausible requirements (for the
absence of singularities) that can, in turn, be used to identify
wide parametric families of potentials with the desired solutions.
Relevant waveforms included, but were arguably not limited to
dark-dark or bright-bright solitary waves and more exotic generalizations
thereof such as the bright wave coupled to a dark-in-bright
structure. Oscillatory variants of such hyperbolic limit
solutions were identified as well.

Naturally, numerous directions of future research arise
from the present considerations. Offering a systematic similar approach
could be of interest also in the case of other nonlinearities. From a stability perspective,
it would appear interesting to identify case examples with
stable isolated parameter
values or, more promisingly, wide parameter ranges, as the
solutions considered here seemed to be largely unstable
(with bands of unstable modes)
resulting in turbulent dynamics 
in many of our dynamical examples.
Finally, exploring two-dimensional generalizations of
the relevant $\cP\cT$-symmetric systems is of particular interest
in its own right both at the level of discrete systems
(see e.g. the plaquette considerations of~\cite{guenther})
and at that of continuum ones (see e.g.~\cite{achilleos}); see also
the recent work of~\cite{yang}.
Such studies are currently in progress and will be reported
in future publications.

\section{Acknowledgments*}
\label{sec:ack}

F.A. acknowledges the support from Grant No. EDW B14-096-0981 provided by IIUM
(Malaysia). P.G.K. gratefully acknowledges the support of 
NSF-DMS-1312856, as well as from
the US-AFOSR under grant FA950-12-1-0332,  
and the ERC under FP7, Marie
Curie Actions, People, International Research Staff
Exchange Scheme (IRSES-605096).

\end{document}